\documentclass[namedate,webpdf,modern,mediumone]{oup-authoring-template}

\usepackage{booktabs,float,tabularx,array,etoolbox}
\graphicspath{{figures/}}
\newcommand{\societylogo}{}
\makeatletter
\patchcmd{\ps@opening}{\color{black!20}\rule{45pt}{55pt}}{}{}{}
\patchcmd{\ps@opening}{\color{black!20}\rule{45pt}{55pt}}{}{}{}
\makeatother
\usepackage[tablesonly,nomarkers,nolists]{endfloat}
\usepackage[section]{placeins}
\usepackage{xurl}
\hypersetup{colorlinks=true,linkcolor=blue,citecolor=blue,urlcolor=blue,
  pdftitle={Scale-dependent contraction of spatial wet-bulb temperature contrasts in eastern China},
  pdfauthor={Jian Hou, Tan Meng, and Maozai Tian}}
\newcolumntype{Y}{>{\raggedright\arraybackslash}X}

\begin{document}

\journaltitle{Journal of the Royal Statistical Society Series C: Applied Statistics}
\copyrightyear{}
\pubyear{}
\lastpage{19}
\title[Spatial wet-bulb contrast contraction]{Scale-dependent contraction of spatial wet-bulb temperature contrasts in eastern China}
\author[1]{Jian Hou\ORCID{0000-0002-9248-6874}}
\author[2]{Tan Meng\ORCID{0009-0001-1812-1834}}
\author[2,$\ast$]{Maozai Tian\ORCID{0000-0002-0515-4477}}
\address[1]{\orgdiv{College of Systems Engineering}, \orgname{National University of Defense Technology}, \orgaddress{Changsha 410073, \country{China}}}
\address[2]{\orgdiv{Center for Applied Statistics, School of Statistics}, \orgname{Renmin University of China}, \orgaddress{\street{59 Zhongguancun Street}, Beijing 100872, \country{China}}}
\corresp[$\ast$]{Corresponding author. \href{mailto:mztian@ruc.edu.cn}{mztian@ruc.edu.cn}}

\abstract{
Regional wet-bulb temperature means omit the spatial distribution of humid
heat. We compare upper-quartile and middle-half days of the monthly regional
mean at 121 sites in a specified eastern-China domain. A prespecified
multiscale architecture combines Gaussian-weighted semivariances at five
bandwidths with equal-month, equal-scale and equal-summer aggregation. The
specification was developed on the 2015 and 2022 summers and applied unchanged
to 33 other summers in 1991--2025. On high-mean days, the weighted mean squared
pairwise WBT contrast was $7.28\%$ lower. This corresponds to a $3.71\%$
reduction after square-root transformation and a $3.97\%$ mean reduction in
record-level root-mean-square pairwise differences. Contraction strengthened
as the Gaussian graph bandwidth increased, from $2.86\%$ at $h=126$ km to
$13.27\%$ at $h=2{,}013$ km. Exact Laplacian decompositions localised the
broad-scale contrast to daily anomalies opposing the monthly north--south
climatological pattern, while anomaly energy changed little. The pattern
persisted in 1950--1990. A sparse external station comparison showed
concordant broad-scale direction. High regional humid heat was associated
with a flatter broad geographic field, a feature that a regional mean cannot
identify.
}


\maketitle

\section{Introduction}

Wet-bulb temperature (WBT) is a widely used measure of humid heat stress that
combines air temperature and humidity \citep{sherwood2010,raymond2020}. Two
days can have identical regional means yet sharply different spatial fields:
one may contain strong north--south or local contrasts, while the other is
comparatively coherent. The analysis therefore focuses on spatial contrast in
the WBT field itself.

Recent work has documented increasingly concentrated humid-heat extremes in
standard climate regions \citep{speizer2022}, sharp local wet-bulb extremes
associated with wet soils \citep{chagnaud2025}, and reduced intra-urban
heat-stress variability during some heat waves \citep{shreevastava2023}.
Spatial-extremes models such as \citet{healy2025} estimate nonstationary
temperature tails and the area exceeding critical thresholds. The within-month
comparison here examines how pairwise WBT differences change with graph
scale. A regional mean discards spatial arrangement, while ordinary cross-site
variance cannot distinguish neighbouring contrasts, mesoscale structure and a
domain-wide gradient. Empirical variograms represent distance-dependent
variation, and kernel smoothing has been used to estimate isotropic
semivariograms nonparametrically \citep{garciasoidan2004}. Here, Gaussian
weights define a multiscale regime-contrast summary for observed fields across
the prespecified distance scales. Smooth weighting provides this summary and
permits an exact spatial allocation of the regional contrast.

We ask whether high regional mean wet-bulb temperature in eastern China is
accompanied by a contraction of spatial differences. The question is
associational. A negative contrast is
compatible with broad atmospheric forcing, but topography \citep{pepin2015},
soil moisture \citep{seneviratne2010}, circulation \citep{horton2015}, and
measurement support can produce or mask similar patterns. If event labels
depend on spatial dispersion as well as the regional mean, the outcome
influences which days are selected, and a difference can be created by
construction \citep{simmons2011}. If the most favourable distance is chosen
after inspecting the data, scale-specific evidence is likewise overstated
\citep{white2000}. Event definitions therefore use regional mean wet-bulb
temperature, and the spatial scales were selected during method development.

The individual statistical ingredients used here are established. Graph
Dirichlet energy is a weighted semivariance under second-order stationarity,
Gaussian weighting is a standard way to smooth over pair distances,
Laplacian projections are familiar, and cyclic shifts use standard
randomisation logic \citep{cressie1993,shuman2013,dong2016}. The contribution
is their prespecified multiscale analysis architecture for this application:
record-specific relative state contrasts, equal aggregation over months,
scales and summers, additive node and structural decompositions, and a joint
shift of each complete daily scale profile. Table~\ref{tab:contribution}
states this boundary explicitly. The architecture reveals which geographic
scales and structures account for a WBT contrast without presenting any one
of its statistical primitives as new.

The application uses 2015 and 2022 for method development and the other 33
summers for evaluation. A product cyclic-shift calculation assesses the
multiscale statistic under a conditional product-invariance null. Equal edge
splitting and Laplacian-orthogonal projections distinguish attenuation of
persistent geographic structure from changes in transient anomalies. The
result matters because two days with the same mean can expose quite different
parts of the domain to humid heat. Broad-scale contraction may be relevant to
the geographic synchrony of threshold exceedance, joint emergency demand and
spatial sampling, but this study does not measure exceedance area, population
exposure or health outcomes.

\begin{table}[H]
\centering
\caption{Boundary of the statistical contribution.}
\label{tab:contribution}
\small
\begin{tabularx}{\textwidth}{@{}p{0.20\textwidth}p{0.24\textwidth}Y@{}}
\toprule
Component & Established basis & Contribution in this study \\
\midrule
Graph Dirichlet energy & Graph-signal roughness & Relative spatial comparison between regional-mean states \\
Kernel-weighted semivariance & Distance-weighted variogram summaries & One hierarchical target across five prespecified bandwidths \\
Cyclic shifts & Group-based randomisation tests & Record-wise joint shifts of complete multiscale daily profiles \\
Node allocation & Edge-contribution splitting & Additivity matched to the relative contrast and aggregation hierarchy \\
Laplacian projection & Graph-orthogonal decomposition & Additive structural localisation of state contrasts \\
Application insight & Not a statistical primitive & Association between high regional humid heat and contraction of the north--south WBT contrast \\
\bottomrule
\end{tabularx}
\end{table}

\section{Multiscale graph regime contrasts}

\subsection{Scale-indexed graph dispersion and regime contrast}

Let $y_{it}$ denote WBT at location $i$ in the selected field for day $t$, and
let $\bar y_t=n^{-1}\sum_i y_{it}$. The benchmark is the site mean
squared spatial deviation,
\begin{equation}
  V_t=\frac{1}{n}\sum_{i=1}^{n}(y_{it}-\bar y_t)^2,
  \qquad n=121.
  \label{eq:variance}
\end{equation}

To retain spatial arrangement, we use an equirectangular kilometre projection
centred at the mean site latitude: longitude is multiplied by
$111.32\cos(\bar\phi)$ and latitude by $110.57$. For
bandwidth $h$, set $w_{ij,h}=\exp\{-d_{ij}^2/(2h^2)\}$ for $i\ne j$ and
$w_{ii,h}=0$. Let $\mathbf W_h=(w_{ij,h})$,
$\mathbf L_h=\operatorname{diag}(\mathbf W_h\mathbf 1)-\mathbf W_h$, and
$S_h=\sum_{i<j}w_{ij,h}$. We define graph dispersion as
\begin{equation}
  Q_h(\mathbf y)=\frac{\mathbf y^\top\mathbf L_h\mathbf y}{2S_h}
  =\frac{\sum_{i<j}w_{ij,h}(y_i-y_j)^2}{2\sum_{i<j}w_{ij,h}}.
  \label{eq:graph-dispersion}
\end{equation}
Graph dispersion is therefore one half of the edge-weighted mean squared
pairwise difference, equivalently an edge-weighted semivariance. Dividing by
$S_h$ makes the measure comparable across graphs with
different total edge weights. Smaller values indicate greater graph-scale
coherence. Graph dispersion measures spatial contrast, distinct from
threshold-exceedance area. It has units $^\circ\mathrm C^2$;
$D_h^{\mathrm{RMS}}=(2Q_h)^{1/2}$ is the weighted root mean squared pairwise
WBT difference in $^\circ\mathrm C$.

Equation~\eqref{eq:graph-dispersion} connects three familiar descriptions of
spatial variation. It is the graph Dirichlet energy used to measure graph
signal roughness \citep{shuman2013,dong2016}. If the field is second-order
stationary with semivariogram
$\gamma(d_{ij})=\tfrac12\operatorname{E}(Y_i-Y_j)^2$, then
\begin{equation}
  \operatorname{E}\{Q_h(\mathbf Y)\}
  =\frac{\sum_{i<j}w_{ij,h}\gamma(d_{ij})}{S_h},
  \label{eq:variogram-average}
\end{equation}
so its expected profile is a kernel average of the semivariogram. Under
second-order stationarity, this identity supplies a semivariogram
interpretation. For the nonstationary WBT fields studied here, $Q_h$ serves as
a graph-scale dispersion measure. On a complete graph with equal off-diagonal
weights, $Q_h=nV/(n-1)$. Ordinary spatial variance is therefore the
complete-graph endpoint of the same family, up to its degrees-of-freedom
factor.

We use five bandwidths selected during method development:
$h/m\in\{0.125,0.25,0.5,1,2\}$, where $m$ is the median pairwise distance.
They correspond to 126, 252, 503, 1,006 and 2,013 km. Their weighted mean pair
distances are 200, 319, 549, 826 and 992 km, and their effective edge counts
are 366, 1,138, 3,223, 6,005 and 7,109 out of 7,260. Thus 126 km denotes a
Gaussian bandwidth with no hard distance cut-off; it is local only relative to
the sampled support, where the median nearest-neighbour distance is 150 km.
For edge weights $w_e$, the reported Kish quantity is
$N_{\mathrm{eff}}=(\sum_e w_e)^2/\sum_e w_e^2$.

Classifying days separately within each month-year record $r=(y,m)$ prevents
the seasonal cycle or an interannual mean shift from entering the event
definition. Let $q_{25,r}$ and $q_{75,r}$ be the type-7 sample quartiles of
$\bar y_t$ in record $r$. A day is labelled high if
$\bar y_t\ge q_{75,r}$ and middle if
$q_{25,r}\le\bar y_t<q_{75,r}$. The analysis excludes lower-quartile days.
There were no threshold ties; every month-year contains eight high days and
14 or 15 middle days. These are relative upper-quartile days: the threshold is
redefined within each record, and no extreme-value tail is involved.

The two groups are compared at each scale by the relative difference of their
dispersion means. Let $\mathcal E_{ym}$ and $\mathcal M_{ym}$ be the high- and
middle-day sets in record $(y,m)$, and let $\overline Q^{E}_{ymh}$ and
$\overline Q^{M}_{ymh}$ denote the corresponding graph-dispersion means.
The scale-specific relative contrast and its summer average are
\begin{equation}
 R_{ymh}=\frac{\overline Q^{E}_{ymh}}{\overline Q^{M}_{ymh}}-1,
 \qquad
 \widehat\theta_y=\frac{1}{3H}\sum_{m=1}^{3}\sum_{h=1}^{H}R_{ymh}.
 \label{eq:relative}
\end{equation}
For a set $\mathcal Y$ of summers, the finite-record target is
\begin{equation}
 \widehat\theta_{\mathcal Y}
 =\frac{1}{|\mathcal Y|}\sum_{y\in\mathcal Y}\widehat\theta_y.
\label{eq:record-estimand}
\end{equation}
A negative value means that high-regional-mean fields vary less across space.
Because the bandwidths are equally spaced on the log scale, the arithmetic
mean gives equal weight to every bandwidth in the prespecified five-scale
grid. It is a protocol-defined finite-record summary, not a unique or
scale-free regional contraction index. The raw ratio is retained as the
protocol-defined primary summary. It gives a direct relative comparison but
is asymmetric, can be sensitive to a small middle-day denominator, and forms
a mean of record ratios rather than a ratio of pooled means. The
log-ratio analysis replaces $R_{ymh}$ by
$\log(\overline Q^{E}_{ymh}/\overline Q^{M}_{ymh})$ at the same three stages
of averaging and back-transforms the final mean for interpretation. A second
analysis uses the bounded symmetric contrast
$2(\overline Q^{E}_{ymh}-\overline Q^{M}_{ymh})/
(\overline Q^{E}_{ymh}+\overline Q^{M}_{ymh})$. The three contrast measures are
applied to identical fields and labels in the denominator stress test. We
treat the log ratio as the principal robustness analysis and the bounded
symmetric form as a secondary check.

\subsection{Exact spatial and structural decomposition}

Graph dispersion accumulates over edges. We obtain an exact and symmetric site
allocation by splitting every undirected edge equally between its endpoints,
although other allocations are possible. For location $i$, define
\begin{equation}
 q_{ih}(\mathbf y)=\frac{1}{4S_h}\sum_{j\ne i}
 w_{ij,h}(y_i-y_j)^2.
 \label{eq:node-energy}
\end{equation}
The identity is exact because the double sum counts every edge twice:
\begin{equation}
 \sum_{i=1}^n q_{ih}(\mathbf y)
 =\frac{1}{2S_h}\sum_{i<j}w_{ij,h}(y_i-y_j)^2
 =Q_h(\mathbf y).
 \label{eq:node-sum}
\end{equation}
The allocation is translation invariant and nonnegative for a single field.
It also respects the graph geometry: a site receives contributions only from
its incident weighted contrasts, while distant pairs carry little weight in a
local graph. Equal edge splitting avoids assigning an arbitrary direction to
an undirected spatial relationship.

Let $\overline q^{E}_{iymh}$ and $\overline q^{M}_{iymh}$ denote the high- and
middle-day means within record $(y,m)$, and let
$\overline Q^{M}_{ymh}=\sum_i\overline q^{M}_{iymh}$. The site contribution to
the record-level relative contrast is
\begin{equation}
 c_{iymh}=\frac{\overline q^{E}_{iymh}-\overline q^{M}_{iymh}}
 {\overline Q^{M}_{ymh}},
 \qquad \sum_i c_{iymh}=R_{ymh}.
 \label{eq:node-contribution}
\end{equation}
The regime-change quantity $c_{iymh}$ may be negative. Its denominator is the
record-specific middle-day
dispersion. Sites in months with different baseline variability therefore
enter the final allocation on the same relative scale as the corresponding
record-level estimator.

Let $\mathcal Y$ be the set of summers. With $H=5$, the reported node
allocation is
\begin{equation}
 c_i=\frac{1}{|\mathcal Y|}\sum_{y\in\mathcal Y}
 \frac{1}{3}\sum_{m=1}^{3}
 \frac{1}{H}\sum_{h=1}^{H}c_{iymh},
 \qquad
 \sum_{i=1}^{n}c_i=\widehat\theta_{\mathcal Y}.
 \label{eq:aggregated-allocation}
\end{equation}
The order and weights in Equation~\eqref{eq:aggregated-allocation} match the
estimator: equal months within a summer, equal graph scales, and equal
summers. The identity is checked numerically after every analysis. Changing
the spatial support changes the incident edges and hence the allocation.
Accordingly, every main-text map is calculated from the 121-node graph. For
a scale-specific map, retaining one $h$ in
Equation~\eqref{eq:aggregated-allocation} yields the corresponding
scale-specific contrast through the observed-node sum.

For display, we fit a low-rank thin-plate REML surface to each set of 121
observed-node values and evaluate it on a fine land-masked raster. Graph
construction, estimation, uncertainty calculations and node-sum identities
all use the observed nodes, which are overlaid on each map. A negative $c_i$
allocates part of the regional contraction to edges incident to site $i$; the
allocation is descriptive at the coordinate level. The maps also show regime
means of $y_{it}-\bar y_t$ and their difference. These spatially centred fields
separate a change in spatial pattern from a shift in the daily regional mean.

To quantify the dominant north--south structure, let $\boldsymbol\ell$ be
centred latitude and $\mathbf z_t=\mathbf y_t-\bar y_t\mathbf 1$. Define
\[
 \widehat\beta_{th}=
 \frac{\boldsymbol\ell^\top\mathbf L_h\mathbf z_t}
      {\boldsymbol\ell^\top\mathbf L_h\boldsymbol\ell},
 \qquad
 \mathbf e_{th}=\mathbf z_t-\widehat\beta_{th}\boldsymbol\ell.
\]
Orthogonality in the $\mathbf L_h$ inner product gives
\begin{equation}
 Q_h(\mathbf y_t)=
 \frac{\widehat\beta_{th}^2\boldsymbol\ell^\top\mathbf L_h\boldsymbol\ell}
      {2S_h}
 +\frac{\mathbf e_{th}^\top\mathbf L_h\mathbf e_{th}}{2S_h}.
 \label{eq:gradient-decomposition}
\end{equation}
High-minus-middle means of the two terms, divided by middle-day total
dispersion, add exactly to $R_{ymh}$. This exact energy accounting is
descriptive. More generally,
for a centred spatial basis matrix $\mathbf B$, the Laplacian projection has
coefficients $\widehat{\boldsymbol\beta}_{th}=
(\mathbf B^\top\mathbf L_h\mathbf B)^{-}\mathbf B^\top\mathbf L_h\mathbf z_t$,
where $(\cdot)^{-}$ is a generalised inverse. Centred latitude is the
reference structural summary. Further projections add longitude and then
elevation, obtained by dividing the official invariant ERA5-Land geopotential
field by standard gravity. We compare the total structured and residual
energies of these nested column spaces; no order-specific latitude, longitude
or elevation contribution is assigned.

A second decomposition separates the persistent monthly field from daily
departures. Write
$\mathbf y_t=\boldsymbol\mu_m+\mathbf a_t$, where $\boldsymbol\mu_m$ is the
site-specific 35-summer mean field for month $m$. Then
\begin{equation}
 Q_h(\mathbf y_t)=
 \frac{\boldsymbol\mu_m^\top\mathbf L_h\boldsymbol\mu_m}{2S_h}
 +\frac{\boldsymbol\mu_m^\top\mathbf L_h\mathbf a_t}{S_h}
 +\frac{\mathbf a_t^\top\mathbf L_h\mathbf a_t}{2S_h}.
 \label{eq:climatology-anomaly}
\end{equation}
The first term is constant within a month and cancels from the high-minus-middle
numerator. The other two terms quantify change in anomaly energy and change in
climatology--anomaly alignment. Dividing both differences by the middle-day
total dispersion yields two components that add exactly to $R_{ymh}$.
The cross term can be negative and is not an energy component. This is an
algebraic, descriptive decomposition conditional on the chosen climatology;
it does not identify circulation, soil-moisture or other physical mechanisms.

\section{Finite-record inference and study design}

\subsection{Product cyclic-shift randomisation}

Days within a month are serially dependent, so each month-year record is
randomised as a unit. Its complete daily outcome profile is rotated by a
randomly chosen offset while the mean-WBT series and labels remain in place. This
joint rotation preserves dependence between bandwidths, and the statistic is
recomputed after every shift. The one-sided Monte Carlo $p$-value is
\begin{equation}
  p=\frac{1+\sum_{b=1}^{B}
  I(T_b^{\mathrm{shift}}\le T_{\mathrm{obs}})}{B+1}.
  \label{eq:mc-p}
\end{equation}

Validity requires cyclic invariance, a stronger property than generic
stationarity. For record $r$,
let $X_r=(\bar y_{rt})_{t=1}^{n_r}$ denote the mean series and let $D_r$
collect the $H$ graph-dispersion values for each day, with
$X=(X_1,\ldots,X_R)$ and $D=(D_1,\ldots,D_R)$. The operator $C_{n_r}$ rotates
all $H$ columns of $D_r$ together, and $G=\prod_r C_{n_r}$ is the group of
all joint rotations. The null hypothesis is group invariance:
\begin{equation}
  (D\mid X)\overset{d}{=}(gD\mid X)\quad\text{for every }g\in G.
  \label{eq:cyclic-null}
\end{equation}
In non-technical terms, conditional on the observed regional-mean series,
independently changing the circular phase of the complete multiscale
dispersion profile in any month-year record must leave the joint distribution
of all profiles unchanged. Under this null, the group-rank argument gives a
finite-sample-valid lower-tail test; the proof and the plus-one Monte Carlo
result are in Supplementary Section~S1 \citep{phipson2010}.

Truncated time-shift tests replace wrap-around invariance with strict
stationarity \citep{harris2020,yuan2024}, but their conservative finite-sample
form requires at least 19 shifts on each side for a 0.05 cutoff. Records of 30
or 31 days leave no usable aligned window under that requirement, so we use
cyclic shifts under Equation~\eqref{eq:cyclic-null}.

The product condition is stronger than marginal cyclic invariance of each
record because it permits the records to be rotated independently. It is also
stronger than a null of zero mean contrast and is not a general test of no
association between regional level and spatial dispersion. Common circulation
or soil-moisture states across adjacent months, seasonal progression and
large-scale climate drivers can violate independent phase invariance. A
targeted simulation in Supplementary Section~S11.4 studies one shared
cross-record process, but it cannot establish the assumption for the observed
climate record. Scale-wise weak family-wise error control and the development-
record summaries are also detailed in Supplementary Section~S1.

\subsection{Development and evaluation design}

The summers of 2015 and 2022 served as the design sample; the other 33 summers
form the evaluation record. Project records show that these were the only
complete summers used during method development, but do not document why
those calendar years were first acquired. We therefore make no claim that
they were randomly sampled or representative. On those summers we inspected WBT fields and graph profiles
and fixed field selection, labels, spatial support, bandwidths, aggregation
and quality control. Event labels, bandwidths and field selection from that
stage carry into Equations~\eqref{eq:relative}--
\eqref{eq:record-estimand}. The target $\widehat\theta_{\mathcal Y_H}$
describes the evaluation record directly; no design-summer result enters it.

For the global cyclic-shift analysis, set
$T_{\mathrm{obs}}=\widehat\theta_{\mathcal Y_H}$. Each Monte Carlo
transformation rotates the five-column dispersion profile within each of the
99 evaluation month-year records, then recomputes every record ratio and the
equal-month, equal-scale and equal-summer mean, with $B=99{,}999$.

We added the global product cyclic-shift distribution after estimating the
33-summer contrast and therefore report its $p$-value as exploratory. The
statistic is the evaluation-record target in
Equation~\eqref{eq:record-estimand}. Its interpretation is restricted to the
conditional product-invariance null in Equation~\eqref{eq:cyclic-null}.

Contrast magnitude is reported through the record mean, summer-specific contrasts,
negative-summer count and leave-one-summer-out range. Supplementary Table~S1
summarises the analysis sequence and the role of each calculation.

The archived Version~2 protocol states a freeze date of 2 August 2026, before
the multi-year point panel was acquired or inspected. Because the first public
repository commit is later, Supplementary Table~S1 records the full provenance
and its evidential limit. Table~\ref{tab:analysis-status} reconciles
the protocol roles, analysis timing and final reporting. The protocol-defined
Student, lag-2 Newey--West and sign-test components, and their three-component
rule, remain confirmatory results in the historical record even where their
final interpretation is more cautious.

\begin{table}[H]
\centering
\caption{Analysis status, timing and reconciliation with the Version~2 protocol.}
\label{tab:analysis-status}
\scriptsize
\setlength{\tabcolsep}{3pt}
\begin{tabularx}{\textwidth}{@{}p{0.17\textwidth}p{0.18\textwidth}p{0.13\textwidth}p{0.20\textwidth}Y@{}}
\toprule
Analysis & Original protocol status & Fixed before the 33-summer result? & Final manuscript role & Reason for final role \\
\midrule
33-summer, five-scale mean & Primary estimand & Yes & Primary finite-record summary & Protocol specification retained \\
Student mean test & Confirmatory component & Yes & Protocol result and process reference & Dependence simulations show under-coverage \\
Lag-2 HAC mean test & Confirmatory component & Yes & Protocol result and process reference & A fixed lag is not generally consistent \\
Independent-summer sign reference & Confirmatory component & Yes & Protocol recurrence reference & Exact binomial interpretation requires independent summer signs \\
Three-component rule & Confirmatory decision rule & Yes & Transparently reported protocol decision & Rule was satisfied \\
Global product shift & Not prespecified & No & Exploratory conditional randomisation & Added after the primary estimate \\
1950--1990 analysis & Pre-access extension & Before historical data access & Temporal extension & Extension specification retained \\
NOAA comparison & Pre-access extension & Before station-result access & External measurement comparison & Uses reanalysis-defined events \\
\bottomrule
\end{tabularx}
\end{table}

\subsection{Process-level reference}

For comparison, a process-level reference treats the summer sequence as a
stationary weakly dependent process. Under standard strong-mixing, moment and
long-run variance conditions, its mean admits a central-limit approximation; a
growing-bandwidth HAC estimator needs a stronger uniform autocovariance condition
\citep{ibragimov1971,newey1987,andrews1991}. The full statement, ratio
negative-moment conditions and proof are in the Supplement. At 33 summers,
simulation shows that Student and HAC intervals can have substantial
undercoverage under serial dependence. Both intervals are reported alongside
the finite-record randomisation result.

\section{Simulation study}

The simulations examine day-level cyclic randomisation and the behaviour of
process-model intervals at the 33-summer record length. The number of summers
determines the repeated-summer sample size, whereas day-level simulations
address within-record randomisation. Complete data-generating specifications
and result tables are in Supplementary Section~S11.

\subsection{Data-generating processes}

The day-level experiment uses the observed coordinates and six 30--31-day
records. Its circular Gaussian null satisfies the invariance condition in
Equation~\eqref{eq:cyclic-null}; two
finite-autoregression variants deliberately violate wrap-around invariance.
Each Gaussian anomaly is projected through
$\mathbf I-\mathbf1\mathbf1^\top/n$, so its sample mean across the 121 sites
is exactly zero and the field mean equals the series used to form labels. The
centred anomaly is independent of that mean. This calibration design isolates
the randomisation properties; the fixed-hour and anomaly-field analyses in
Section~5 examine field selection and label--field dependence in the observed
record.
Amplitude reduction, correlation-range expansion and gradient suppression
produce true profile contractions of 5.9\%--27.4\%, calculated from the known
quadratic forms. The Monte Carlo standard error is 0.7 percentage points at
5\% rejection and 1.6 points at 50\% power, so differences of one or two
points are difficult to distinguish.

The comparison includes graph and binned-variogram profiles, spatial
variance, two local semivariances, Moran's $I$ and Geary's $C$. These are a
finite set of deliberately constructed mechanisms, not a claim of general
optimality. The repeated-summer experiment compares Student, fixed-lag and
growing-lag calculations, the sign test and the protocol's three-component
rule.

\subsection{Calibration and comparison across spatial mechanisms}

Figure~\ref{fig:simulation} summarises the power ranges over nine spatial
alternatives; complete values are in the Supplement.

Across eight cyclic-invariant designs, graph-profile rejection ranged from
4.5\% to 6.8\%, while the six competitors spanned 4.0\%--6.4\%.
The graph test had rejection rates of 4.5\%
at temporal correlation 0.75 and 6.1\% under the $t_3$ field. The two
non-circular boundary cases gave 4.6\% and 5.1\%.

Gaussian-alternative absolute bias was below 0.48 percentage points and root
mean squared error was 2.6--3.8 points. Under the original $t_3$ null, graph
randomisation size was 6.1\%. A paired denominator stress test then used
2,000 heavy-tailed data sets under each of the null and $-7\%$ alternatives
and 999 common product shifts per data set. Null rejection was 4.70\%, 4.90\%
and 4.75\% for the raw ratio, log ratio and bounded symmetric contrast. On
their native scales, the corresponding null absolute biases were 0.372, 0.070
and 0.066, and root mean squared errors were 1.472, 0.328 and 0.277. Under the
$-7\%$ alternative, root mean squared errors were 1.369, 0.328 and 0.275, while
rejection was 7.95\%, 8.50\% and 7.90\%. The bounded contrast had sharply
lower estimator dispersion, with similar power in this design. Randomisation
calibration can coexist with unstable effect estimation; resistant summaries
reduce the instability while leaving residual estimator variation.

Performance varies across the nine simulated mechanisms.
Nearest-neighbour semivariance is strongest for range changes, but its power is 40.8\%
under weak gradient suppression, compared with 90.9\% for the graph profile
and 96.5\% for variance. The five-bin variogram is similar to the graph profile
for gradient suppression (93.9\% versus 90.9\% at the weakest setting), but is
less sensitive to the weak amplitude and range changes (33.6\% and 19.0\%
versus 50.0\% and 48.9\%). Moran's $I$ and Geary's $C$ have near-nominal power
for pure amplitude changes because standardisation removes amplitude. The
profile's minimum power across the nine alternatives was 48.9\%, higher than
the minimum of each comparator within this experiment only.

The scale-specific profiles retain information lost by a scalar test. Strong
amplitude reduction changes all three summaries by about 19\%; range expansion
acts most strongly on the local graph (28.2\%), while gradient suppression
acts on variance and the broad graph (43.9\% and 42.5\%) and changes the local
graph by 9.3\%.

\begin{figure}[!t]
\centering
  \includegraphics[width=\textwidth, height=0.62\textheight, keepaspectratio]{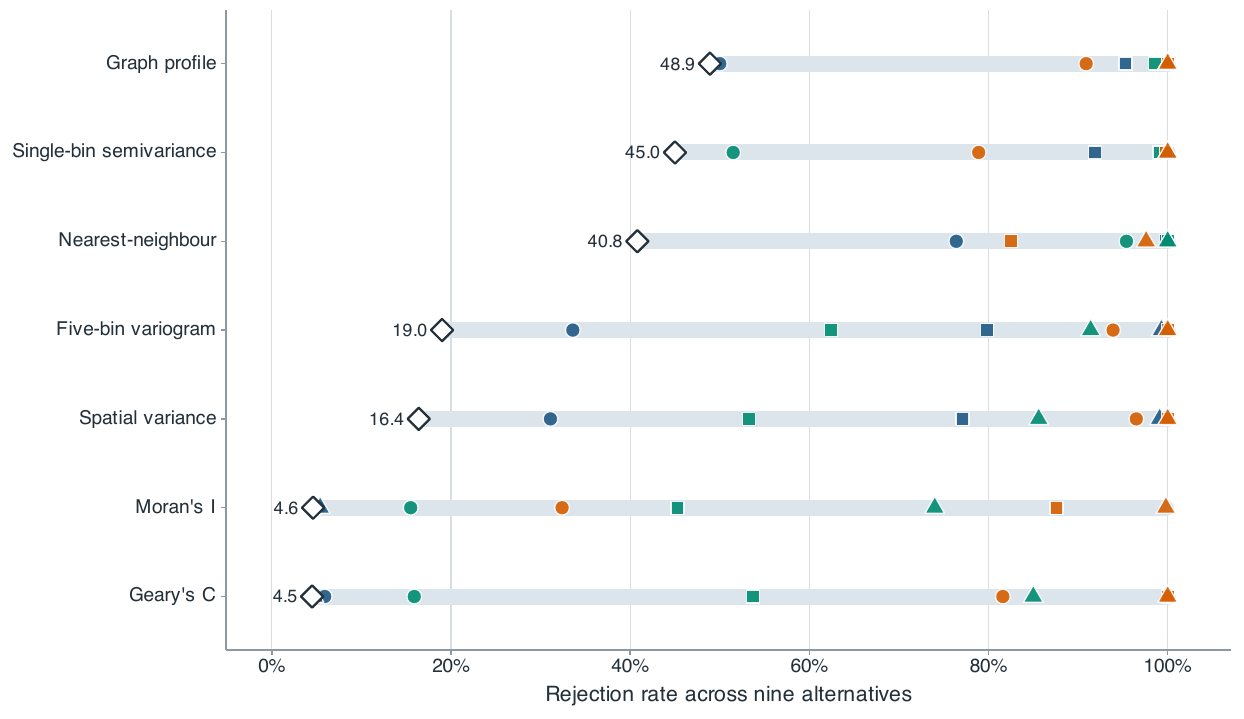}
  \caption{Power across nine simulated spatial alternatives. Blue, green and
  orange marks denote amplitude reduction, correlation-range expansion and
  gradient suppression; circles, squares and triangles denote weak, medium
  and strong changes. For each method, the pale segment spans power across the
  nine alternatives and the diamond marks the minimum.}
  \figalttext{A range-and-dot plot compares seven methods.
  Each method has nine coloured and shaped points, a pale segment spanning
  their range and a diamond at the minimum; the ranges overlap substantially.}
  \label{fig:simulation}
\end{figure}

Eight additional scenarios allow the label-generating mean and spatial field
to share structure; full specifications are in the Supplement. In
the shared-latent null, the sign of the spatial
gradient has mean correlation 0.61 with regional WBT, but its squared energy
is constant, so the true dispersion contrast is zero. The other nulls add a
within-month trend, anisotropic spatial covariance or selection of a daily
peak from 24 simulated hourly regional means, with hour-invariant anomaly
covariance. Corresponding alternatives attenuate the gradient, impose a
seasonally declining gradient or reduce high-day anomaly amplitude. Graph
null rejection is 4.0\%--7.0\%, compared with 4.3\%--7.3\% for the five-bin
variogram. Graph rejection under these four nulls is close to its
cyclic-invariance calibration. The graph profile is at least as
powerful in three of four alternatives, whereas the variogram is slightly
stronger for the seasonal-gradient alternative. Smooth graph weighting and
exact allocation therefore show no consistent power penalty across these
scenarios.

\subsection{Process-reference stress test}

At $R=33$, increasing Gaussian year correlation from 0 to 0.6 reduces Student
coverage from 95.2\% to 66.1\%, fixed-lag coverage from 93.3\% to 78.7\%, and
growing-HAC coverage from 91.2\% to 79.0\%. Corresponding Gaussian null
rejection rises to 21.2\%, 14.4\% and 14.1\%. The error arises mainly from
long-run variance estimation. Coverage from all three process-model intervals
deteriorates at this record length under the simulated dependence. Full
results for 108 cells, including skewed and heavy-tailed
innovations, appear in the Supplement.

An additional 10,000-replication experiment uses the exact 1991--2025 calendar
and removes 2015 and 2022. The largest absolute mean bias over correlations
0, 0.3 and 0.6 and effects 0 and $-7\%$ is 0.03 percentage points. Excluding
the two design summers therefore has negligible first-order centring bias,
but short-record dependence remains: at correlation 0.6, null coverage is
68.3\% for the Student interval that treats summers as independent and 80.2\%
for calendar lag 2.

\section{Humid-heat application}

We apply graph dispersion to synchronous wet-bulb temperature fields in
eastern China during 1991--2025 and extend the analysis to 1950--1990.

\subsection{Data construction and analysis design}

ERA5-Land supplies hourly land-surface fields on a $0.1^\circ$
latitude--longitude grid \citep{munoz2021}. We use 2-m air temperature, 2-m
dew-point temperature and surface pressure for June--August 1991--2025 over
105--125$^\circ$E and 20--42$^\circ$N. The rectangle provides
a regular support spanning the subtropical-to-temperate north--south gradient
of eastern China while excluding most of the Tibetan Plateau west of
105$^\circ$E. Its northern and southern limits bound that gradient. The
analysis panel begins in 1991.

The historical extension covers all 41 June--August seasons from 1950, the
first year of the ERA5-Land record, through 1990. It uses the same sites,
field construction and five-scale estimator. We report separate summaries for
1950--1978 and 1979--1990 because the earlier segment uses the preliminary
ERA5 back extension
(\href{https://confluence.ecmwf.int/pages/viewpage.action?pageId=402639006}{ECMWF
production documentation}). The extension plan and its 99,999-draw product
cyclic-shift calculation were recorded before the historical values were
retrieved.

\begin{figure}[!t]
\centering
  \includegraphics[width=0.90\textwidth, height=0.60\textheight, keepaspectratio]{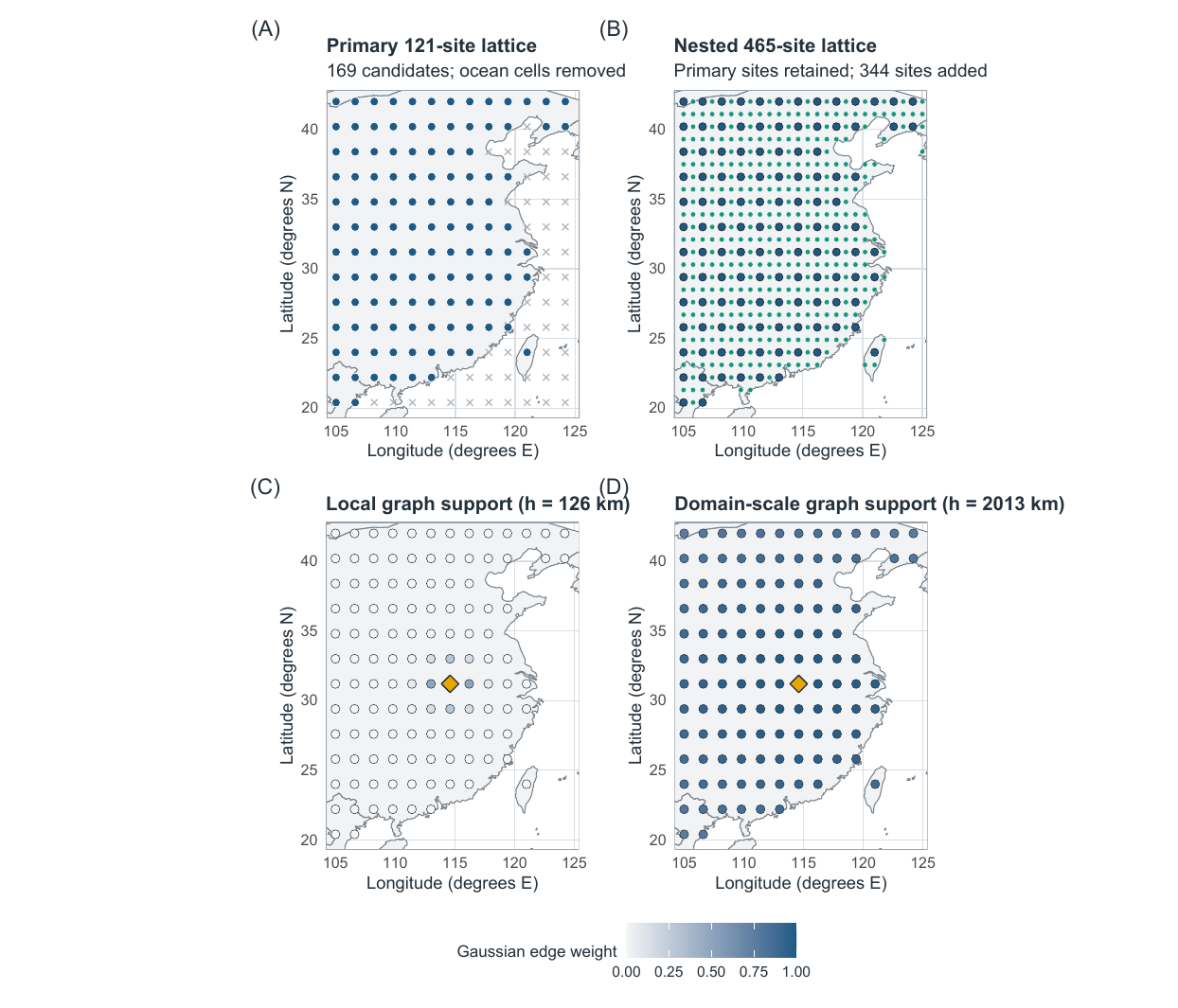}
  \caption{Spatial sampling and graph support. (A) Analysis lattice with 121
  land sites from 169 candidates. (B) Nested lattice adding 344 land sites
  while retaining every analysis site. (C,D) Gaussian edge weights from one
  reference site under the 126- and 2,013-km graphs, showing how bandwidth
  changes the geographic support at fixed coordinates. Basemap: Natural Earth
  1:50m, accessed through the R \texttt{maps} package.}
  \figalttext{Four eastern-China maps compare the analysis and
  nested site lattices, then show weights concentrated near one reference site
  under the 126-km graph and spread across the domain under the 2,013-km graph.}
  \label{fig:study-area}
\end{figure}

Every 16th longitude and 18th latitude cell from a fixed north-west origin
gives a $13\times13$ candidate lattice; removing ocean cells leaves 121 sites
(Figure~\ref{fig:study-area}). The 35 summers contain 9,350,880
quality-controlled hour--site records and 92 complete UTC fields per summer.
Because of NetCDF packing, 2,092 dew-point values exceeded air temperature
slightly (0.022\% of records); these pairs were clipped at equality before WBT
was calculated.
The main target is a site average. For area weighting,
$a_i=\cos(\mathrm{latitude}_i)$,
\[
 \bar y_t^{(a)}=\frac{\sum_i a_i y_{it}}{\sum_i a_i},\qquad
 Q_{h,a}(\mathbf y)=
 \frac{\sum_{i<j}a_i a_j w_{ij,h}(y_i-y_j)^2}
      {2\sum_{i<j}a_i a_j w_{ij,h}}.
\]
The first area-weighted calculation replaces $Q_h$ by $Q_{h,a}$ using the
original labels; the second redefines the labels from $\bar y_t^{(a)}$ at the
regional peak hours.
The two calculations separate spatial weighting from event relabelling.

WBT is calculated from temperature $T$ and dew point $T_d$ in kelvin and
surface pressure $p$ in pascals. Vapour pressure is
$e(T_d)=611.2\exp\{17.67(T_d-273.15)/(T_d-29.65)\}$ Pa, and the Bolton
mixing-ratio, lifting-condensation-temperature and equivalent-potential-
temperature relations are evaluated in these units \citep{bolton1980}.
Saturated $T_w$ is obtained by solving
\begin{equation}
  \theta_e(T_w,T_w,p)=\theta_e(T,T_d,p)
  \label{eq:wbt}
\end{equation}
at the original pressure with 40 bisection iterations on
$[T_d,T]$; even a 100-K initial interval then falls below $10^{-9}$ K. Values
with $T_d>T$ are clipped at equality before evaluation. The Stull approximation
provides a comparison \citep{stull2011}; the main calculation follows the
pseudoadiabatic definition discussed by \citet{davies2008}.

For each UTC day, the 121 simultaneous values at the peak regional mean form
the daily field. Within each month-year, the upper regional-mean quartile is
high, the middle half is middle and the lower quartile is excluded. Labels
depend on regional mean WBT, and the selected hour maximises that mean; graph
dispersion enters after selection. UTC+8 days,
alternative thresholds, sitewise maxima, all 24 fixed UTC hours and daily-mean
fields provide comparisons. At each fixed hour, both the original
peak-based labels and labels recomputed from that hour's regional mean are
analysed; daily-mean labels are likewise recomputed from daily-mean WBT\@. A
dated event manifest records the type-7 thresholds, membership and absence of
ties.

Three checks address within-month seasonal progression: comparison of the
day-of-month distributions, removal of a separate
linear day-of-month trend at every site within each year--month, and
subtraction of each site's calendar-day climatology, estimated from all
summers except the one being analysed. The last two calculations use the
original event labels.
A date-matched calculation compares each high day with
all middle days in the same month-year lying within $\pm3$ or $\pm5$ calendar
days. It first averages the available high-day-specific ratios and then uses
the same equal-month, equal-scale and equal-summer hierarchy as the
estimator; unmatched-day and empty-record counts are retained.

A continuous diagnostic regresses $\log Q_h$ on centred regional
mean WBT within each month-year and bandwidth. Slopes are then averaged equally
over months, bandwidths and evaluation summers. This slope supplements the
quartile target with a graded summary.

Alternative graph kernels are $\exp(-d_{ij}/\rho)$ and
$[1-d_{ij}/\rho]_+^2$. For each Gaussian scale, $\rho$ is found by bisection so
that the alternative kernel has the same edge-weighted mean pair distance.
Across the five scales, their effective edge counts range from 334 to 7,112
and from 382 to 7,113, compared with 366 to 7,109 for the Gaussian graphs;
the full matched-distance and effective-edge metadata accompany the analysis.

\begin{figure}[!t]
\centering
  \includegraphics[width=0.90\textwidth, height=0.60\textheight, keepaspectratio]{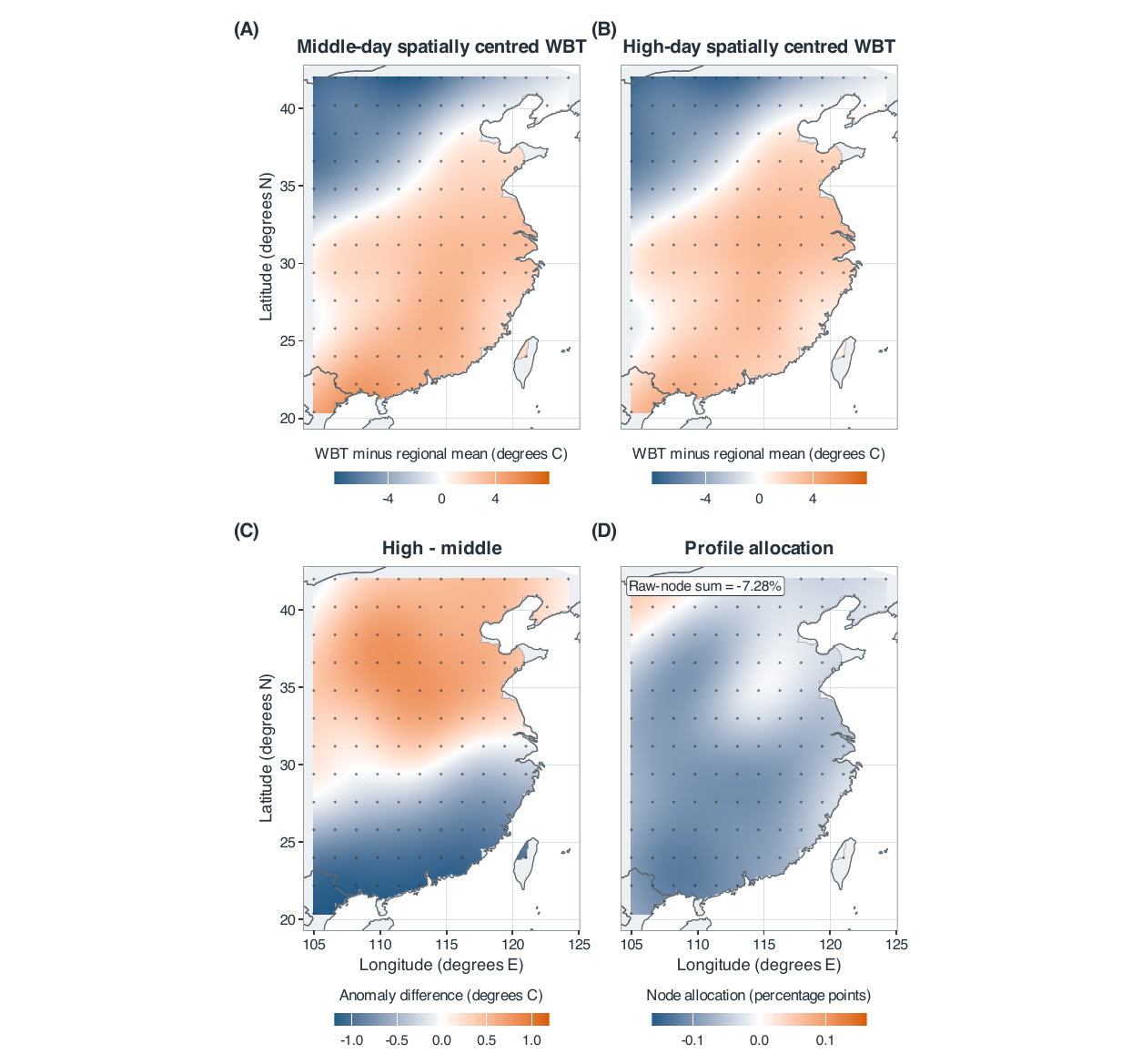}
  \caption{Spatial structure of the 121-site evaluation contrast.
  (A,B) Middle- and high-day spatially centred fields; (C) their difference;
  (D) exact observed-node allocations averaged over the five
  graph scales. The 121 allocations sum to $-7.28\%$, identical to the
  five-scale estimand. The interpolated surfaces visualise the 121
  observed-node values; estimation and inference use the observed nodes.}
  \figalttext{Four smoothed eastern-China maps derived from
  121 analysis sites, shown as dots. The spatially centred surfaces show a weaker
  north--south contrast on high days; the
  high-minus-middle surface changes from positive in the north to negative in
  the south, and most observed-node allocations are negative, with the
  strongest negative values concentrated in the central and southern parts
  of the domain.}
  \label{fig:spatial-decomposition}
\end{figure}

A nested $0.8^\circ\times0.9^\circ$ lattice adds 344 land sites while retaining
all 121 locations. Numerical convergence is assessed at the five physical
bandwidths by comparing 121 sites with a 239-site longitude refinement, a
236-site latitude refinement and all 465 sites under the evaluation events;
peak times and labels are also reselected on the full grid. Thin-plate REML
surfaces \citep{wood2003} are used only
for display in Figures~\ref{fig:multiscale-evidence} and
\ref{fig:spatial-decomposition}; every estimate uses observed nodes.

Additional spatial-support checks use WGS84 geodesic distances on the 121
sites and the 465-site lattice. On the latter, each boundary is moved inward
by one or two lattice steps and event labels are recomputed from the retained
domain. A non-rectangular support intersects the downloaded grid with the
Natural Earth low-resolution China feature; Taiwan is a separate feature in
that source and is not included. Only inward changes are estimable because no
fields were downloaded outside the original rectangle. These checks retain
the five physical bandwidths and report complete scale curves.

We compare ERA5-Land WBT with observations from the National Oceanic and
Atmospheric Administration Integrated Surface Database (NOAA ISD)
\citep{smith2011}. The station comparison uses ten summers (1992, 1996, 2000,
2004, 2008, 2012, 2016,
2020, 2023 and 2025) selected before station retrieval. Archive,
operating-period and geographic eligibility rules followed by a deterministic
maximin design selected 30 dispersed stations. A station-year was retained
when it had
at least 400 quality-controlled, complete exact-hour observations of
temperature, dew point and sea-level pressure; station elevation supplied
surface pressure under a standard atmosphere,
$p_s=p_{\mathrm{SL}}\{1-2.25577\times10^{-5}z\}^{5.2559}$ for elevation $z$
in metres. At two offshore stations, all
variables were missing from the nearest ERA5-Land grid cells, leaving 28
usable stations and 175,172 exact-hour WBT pairs.

The station-field calculation uses peak times and high--middle labels from the
121-site ERA5-Land fields. At each retained time, NOAA and ERA5-Land graph
contrasts use the identical station subset and require at least 10 stations.
All 30 year-month records retained at least one field in each regime. The
503-, 1,006- and 2,013-km graphs were designated as station-supported before
evaluation; the two narrower graphs diagnose the limit of the station
geometry. The comparison assesses measurement agreement and the direction of
spatial contrast on this sparse network. Because times and labels come from
ERA5-Land and station support varies across fields, it is not an independent
event replication.

\subsection{Evaluation-period multiscale result}

Across the 33 evaluation summers, the five-scale contrast on the weighted
squared-difference scale was $-7.28\%$. The protocol-defined three-component
consistency rule was satisfied. The one-sided Student and lag-2 Newey--West
reference values were $2.56\times10^{-6}$ and $3.54\times10^{-6}$; the
independent-summer exact binomial sign reference value was
$3.31\times10^{-5}$. All three were below the prespecified 0.025 threshold.
The corresponding two-sided Student and lag-2 HAC reference
intervals were $[-10.00,-4.57]\%$ and $[-10.05,-4.51]\%$. Subsequent
simulations nevertheless showed that Student and fixed-lag HAC intervals can
under-cover under appreciable interannual dependence. We report the protocol
decision but do not treat it as definitive long-run process inference.

In the post-result exploratory calculation, no sampled shift was as negative
as the observed statistic. With $B=99{,}999$ draws, an exceedance count of
zero and seed 20260809, the plus-one Monte Carlo value was
$p_{\mathrm{MC}}=1.0\times10^{-5}$, the minimum attainable value for this
run. The complete five-scale daily profile was shifted jointly within each of
the 99 month-year records. This result applies only under the conditional
product-invariance null in Equation~\eqref{eq:cyclic-null}.

Twenty-eight summer-specific contrasts were negative, and leave-one-summer-out
estimates ranged from $-7.69\%$ to $-6.64\%$. The Student interval above is a
descriptive $t$-reference interval based on between-summer variation, not a
confidence interval under serial dependence. The fraction negative is 0.85,
with a binomial Wilson reference interval of $[0.69,0.93]$ under independent
summer signs and a one-sided independent-summer sign reference value of
$3.3\times10^{-5}$.

The fitted trend was $-0.22$ percentage points per year (95\% descriptive OLS
interval based on between-summer variation, $[-0.48,0.05]$, two-sided $p=0.10$).
The 2008--2025 mean was 2.95 points more negative than the 1991--2007 mean,
with a descriptive Welch interval based on between-summer variation of
$[-8.43,2.52]$ points. Both intervals include zero,
leaving the temporal trend uncertain.

\begin{figure}[!t]
\centering
  \includegraphics[width=\textwidth, height=0.62\textheight, keepaspectratio]{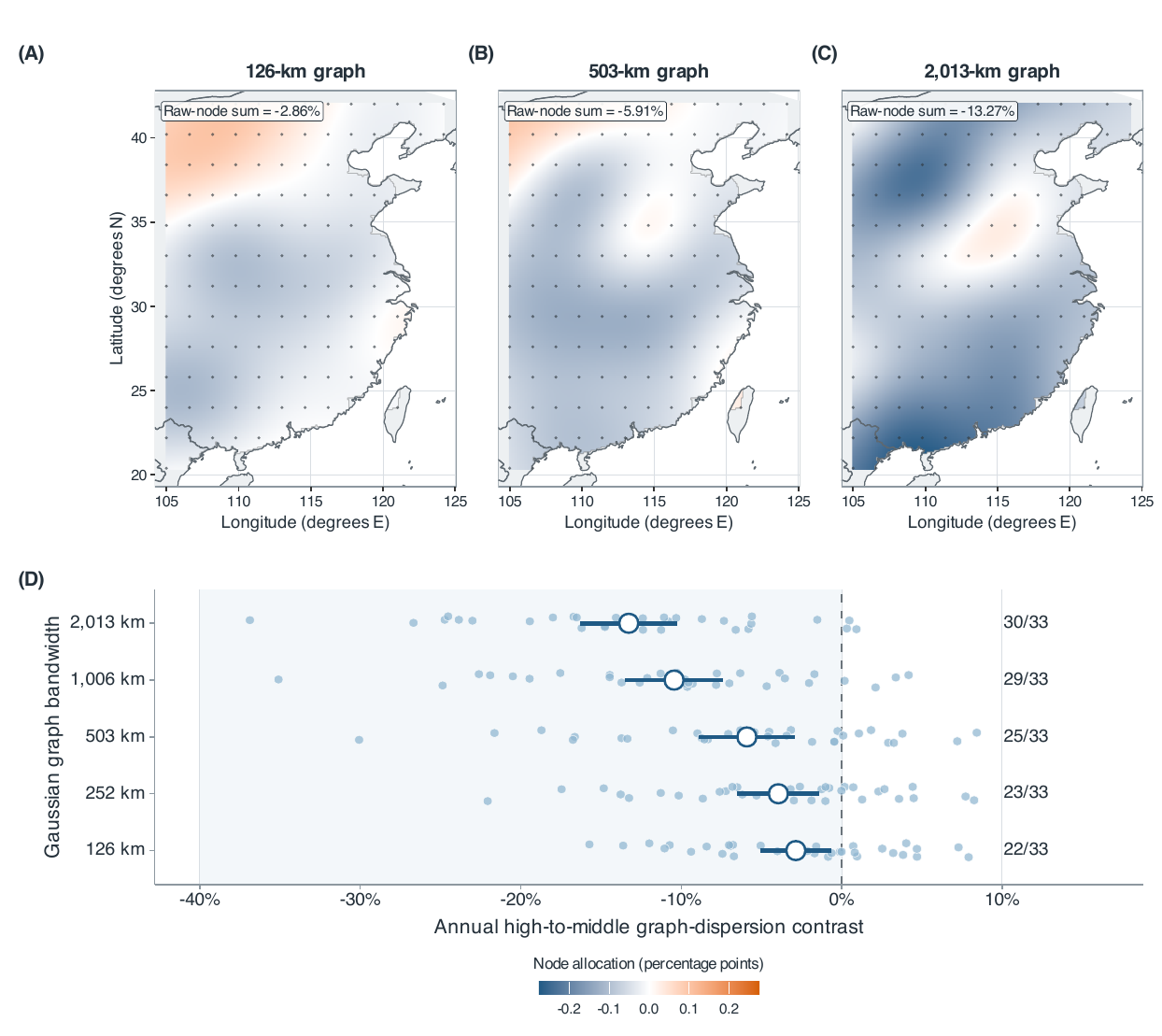}
  \caption{Scale-specific node allocations and summer contrasts for the 33
  evaluation summers. (A--C) Allocations for the 126-, 503- and 2,013-km
  graphs share a symmetric colour scale; their observed-node sums are
  $-2.86\%$, $-5.91\%$ and $-13.27\%$. The interpolated surfaces visualise the
  121 observed-node values; estimation and inference use the observed nodes.
  (D) Small circles show summer-specific contrasts at all five bandwidths;
  large circles and bars show means and 95\% descriptive $(t)$-reference intervals
  based on between-summer variation. Labels
  give the numbers of negative summers.}
  \figalttext{Three smoothed eastern-China maps, each
  overlaid with the 121 analysis sites, show increasingly coherent negative
  node allocations at 126, 503 and 2,013 km. A lower panel shows the 33 annual
  contrasts at all five bandwidths, with mean contrasts and negative-summer
  counts increasing in magnitude with bandwidth.}
  \label{fig:multiscale-evidence}
\end{figure}

Contrasts increase in magnitude from $-2.86\%$ as the Gaussian bandwidth is
$h=126$ km to $-13.27\%$ at $h=2{,}013$ km
(Table~\ref{tab:confirm-scales}; Figure~\ref{fig:multiscale-evidence}).
Negative-summer counts rise from 22 to 30. A 31-point log-spaced bandwidth
profile decreases at each successive evaluated bandwidth across the same
$h=126$--2,013-km range.
Complete-graph variance falls 14.45\% and is negative in all 33
summers.

\subsubsection*{Historical extension, 1950--1990}
The 1950--1990 extension gave a five-scale contrast of
$-11.04\%$ (descriptive $t$-reference interval based on between-summer variation,
$[-12.84,-9.24]\%$). Forty of 41 summer contrasts were negative, and the
leave-one-summer-out range was $-11.33\%$ to $-10.62\%$. Its 99,999-draw
product-shift calculation produced no statistic as negative as the observed
value
(plus-one $p_{\mathrm{MC}}=1.0\times10^{-5}$, the minimum attainable value
for that run, under its conditional product-invariance null). The scale
profile was $-5.96,-7.51,-10.08,-14.50$ and
$-17.17\%$ from 126 to 2,013 km, with 38, 39, 40, 41 and 41 negative summers.
The negative sign and strengthening contraction with bandwidth also appear in
the 41 earlier summers.

The production-history split gave $-10.50\%$ in 1950--1978 (28 of 29 summers
negative) and $-12.34\%$ in 1979--1990 (12 of 12 negative). The segments are
reported separately because the pre-1979 record uses different forcing.
The historical climatology--anomaly decomposition re-estimates the
site-specific monthly climatological field from the 41 summers in 1950--1990.
At 2,013 km, the historical anomaly-energy component was $-0.28$ points
(95\% descriptive $(t)$-reference interval based on between-summer variation,
$[-1.01,0.45]$),
whereas the climatology--anomaly cross component was $-16.89$ points
($[-19.31,-14.47]$) and negative in all 41 summers. Both the
latitude and latitude--longitude structured components were negative in every
summer at every bandwidth. In the earlier record, most broad-scale contraction
was likewise in the alignment term. The 1950--1978 estimates inherit the
uncertainty of the preliminary forcing data.

\subsection{Spatial structure and energy decomposition}

The decomposition in Equation~\eqref{eq:gradient-decomposition}
localises most of the contrast to the north--south gradient. At 126 km it
accounts for $-2.58$ points (95\% $(t)$-reference interval under an
independent-summer model, $[-2.92,-2.24]$) of the $-2.86$-point contrast. From
252 to 2,013 km its mean contribution is more negative than the total
($-4.60,-8.38,-12.73,-14.96$ points), while residual contrasts contribute
$+0.66,+2.48,+2.29,+1.69$ points. The gradient component is negative in all
33 evaluation summers at every bandwidth. Adding longitude to the basis makes
the 2,013-km structured component $-16.23$ points, compared with $-14.96$
points for latitude alone; the corresponding residuals are positive. The
latitude--longitude--elevation basis gives a 2,013-km structured component of
$-14.34$ points (95\% descriptive $(t)$-reference interval based on
between-summer variation, $[-16.75,-11.92]$), negative in all 33 summers, and
a residual of $+1.07$ points. At 126 km, however, its structured component is
$-1.16$ points (95\% descriptive $(t)$-reference interval based on
between-summer variation, $[-2.66,0.33]$) and is negative in only 18 summers.
At broad scales, the
nested bases show that latitude accounts for most of the low-dimensional
geographic--topographic structure; structured localisation is weaker at 126 km. Node
allocations are negative at 106 sites and sum exactly to $-7.28\%$. Each is a
symmetric incident-edge allocation.

The climatology--anomaly decomposition in
Equation~\eqref{eq:climatology-anomaly} provides a descriptive location of the broad-scale change. At
126 km, anomaly-energy change
contributes $-2.63$ points and the climatology--anomaly cross term contributes
$-0.22$ points. At 2,013 km, the anomaly-energy component is $+0.41$ points
(95\% descriptive $(t)$-reference interval based on between-summer variation,
$[-0.50,1.31]$), whereas the cross term is $-13.68$ points
(95\% descriptive $(t)$-reference interval based on between-summer variation,
$[-16.68,-10.67]$) and is negative in all 33 summers. Thus the broad-scale
contraction was accounted for algebraically by high-day anomalies opposing
the chosen monthly climatological pattern, while their own broad-scale energy
changed little. The cross term is not an energy and this accounting does not
identify a physical mechanism.

For each evaluation summer, a sensitivity calculation rebuilt the three
monthly climatological fields from the other 34 summers and repeated the
exact decomposition. At $h=2{,}013$ km, this leave-one-summer-out version gave
$+0.40$ percentage points for anomaly energy and $-13.67$ points for the
cross term, compared with $+0.41$ and $-13.68$ points for the inclusive
climatology. The maximum numerical failure of the daily identity was
$7.2\times10^{-14}$. The structural interpretation is therefore insensitive
to including the evaluated summer in its climatological reference.

\begin{figure}[!t]
\centering
  \includegraphics[width=0.92\textwidth, height=0.60\textheight, keepaspectratio]{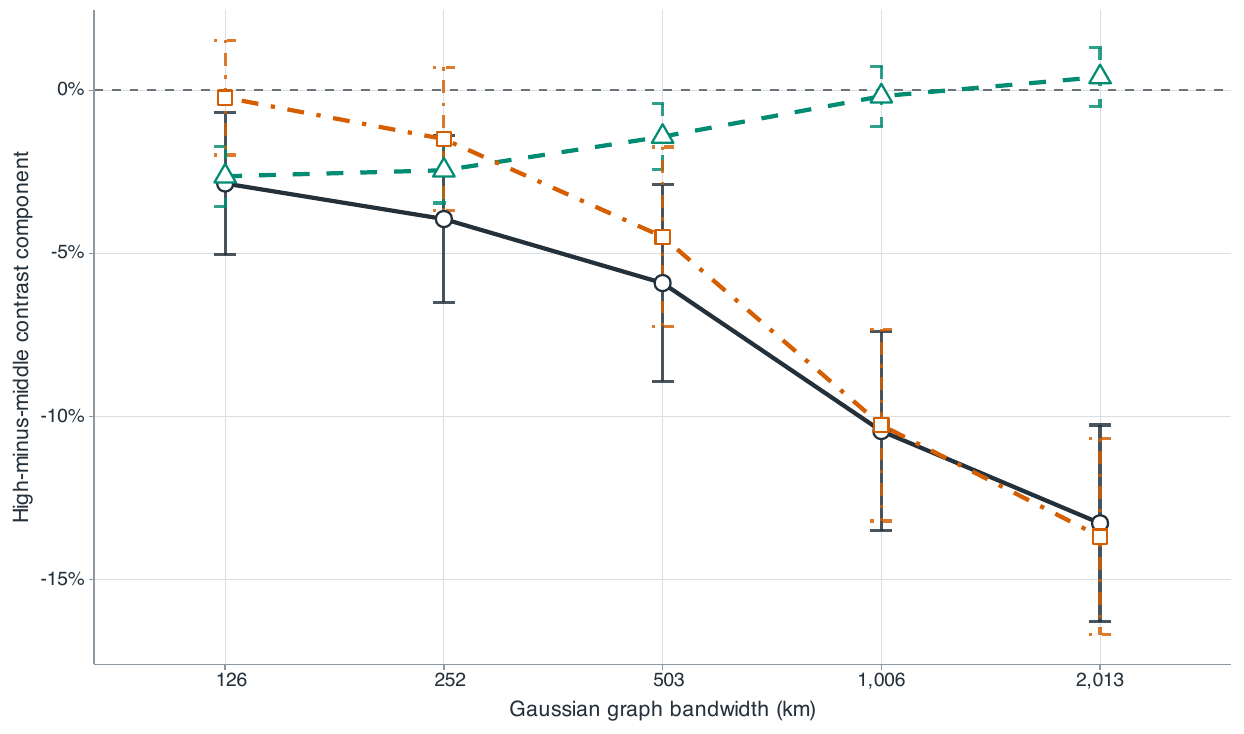}
  \caption{Exact decomposition of the raw-field graph contrast. Dark circles,
  teal triangles and orange squares denote total change, anomaly-energy change
  and the climatology--anomaly cross component; bars are 95\% descriptive
  intervals. Monthly-climatology energy cancels within each record.}
  \figalttext{Across five bandwidths, dark circles become
  increasingly negative. Teal triangles remain near zero at broad scales,
  while orange squares closely follow the negative total.}
  \label{fig:energy-decomposition}
\end{figure}

The convergence analysis uses the primary equirectangular distance matrix and
compares each grid with the 465-site result using the evaluation events. The
121-site scale profile differs by at most 0.99
percentage points, with a five-scale root mean squared difference of 0.76
points. Latitude refinement reduces these values to 0.50 and 0.38 points;
longitude refinement gives 0.93 and 0.73 points. On the 465-site grid, the
primary distances and original events give $-8.03\%$; retaining those
distances but reselecting events gives $-7.25\%$.
Contraction strengthens with bandwidth at every resolution. Agreement between
spatial supports is weakest at 126 km, where the 121-site spacing and station
support are coarsest.

Using the primary equirectangular distance matrix, cosine-latitude weighting
gave $-8.11\%$ with the original labels and $-7.67\%$ after relabelling.
Replacing the distance matrix by WGS84 geodesic distances gave corresponding
estimates of $-7.80\%$ and $-7.37\%$; the WGS84 equal-site estimate was
$-6.97\%$. On the full 465-site domain, WGS84 distances with reselected events
gave $-6.93\%$, and relabelled boundary checks ranged from $-6.32\%$ to
$-7.97\%$. The non-rectangular Natural Earth China-land intersection gave
$-6.35\%$, with a scale curve of $-2.09,-2.80,-5.06,-9.64$ and $-12.18\%$.
Every support retained the negative direction and strengthening contraction
with bandwidth, although the magnitude depends on the specified domain.

\begin{table}[H]
\centering
\caption{Physical scale of the evaluation-period graph contrasts. $h$ is the
Gaussian bandwidth, not a pair-distance cutoff. $\bar d_w$ is the
edge-weighted mean pair distance, and $N_{\mathrm{eff}}$ is the Kish effective
edge count. Graph contrasts are relative changes in weighted squared
differences. RMS contrasts average record-specific ratios of
$(2Q)^{1/2}$; Negative is the number of negative summer graph contrasts.}
\label{tab:confirm-scales}
\small
\setlength{\tabcolsep}{4.5pt}
\begin{tabular}{rrrrrr}
\toprule
$h$ & $\bar d_w$ & $N_{\mathrm{eff}}$ & Graph contrast & RMS contrast & Negative \\
(km) & (km) & (edges) & (\%) & (\%) & (/33) \\
\midrule
126   & 200 & 366   & $-2.86$  & $-1.54$ & 22 \\
252   & 319 & 1,138 & $-3.94$  & $-2.14$ & 23 \\
503   & 549 & 3,223 & $-5.91$  & $-3.25$ & 25 \\
1,006 & 826 & 6,005 & $-10.44$ & $-5.70$ & 29 \\
2,013 & 992 & 7,109 & $-13.27$ & $-7.22$ & 30 \\
\bottomrule
\end{tabular}
\end{table}

The reported $-7.28\%$ is an equally aggregated relative change on the
squared-difference, or semivariance, scale. It corresponds to a $-3.71\%$ RMS
change when transformed as $\sqrt{1-0.0728}-1$; averaging record-specific RMS
ratios gives $-3.97\%$.

Threshold, Stull and sitewise-maximum sensitivities range from $-6.90\%$ to
$-8.27\%$. Early- and late-period summaries have overlapping descriptive
intervals. These checks assess the stability of the spatial contrast across
analysis choices.

For comparison, the same equal-month, equal-scale and equal-year formula gives
$-15.14\%$ for the 2015 and 2022 design summers.

\subsection{Robustness and external measurement agreement}

The 495 middle-day denominators range from 2.25 to
$25.99\;{}^\circ\mathrm{C}^2$ across records and scales; within-scale
coefficients of variation are 0.076--0.177, with a minimum of
$2.25\;{}^\circ\mathrm{C}^2$. These observed denominators do not establish
general estimator stability. In the heavy-tail simulation, randomisation
calibration coexisted with substantially higher raw-ratio RMSE. The
back-transformed log-ratio, our principal robustness result, is $-8.30\%$; the bounded
symmetric contrast $2(Q^H-Q^M)/(Q^H+Q^M)$ is $-8.59\%$. With the primary
equirectangular distances, cosine-latitude weighting gives $-8.11\%$ using the
original labels and $-7.67\%$ after relabelling. With WGS84 distances, the
corresponding estimates are $-7.80\%$ and $-7.37\%$. Kernels matched to the Gaussian weighted
mean distance give $-7.71\%$ (exponential) and $-7.33\%$ (compact quadratic).
All preserve the negative direction and strengthening contraction with
bandwidth.

The continuous-intensity diagnostic gives a five-scale $\log Q_h$ slope of
$-0.065\;{}^\circ\mathrm C^{-1}$ (95\% descriptive $(t)$-reference interval
based on between-summer variation,
$[-0.082,-0.048]$), with negative summer slopes in 31 of
33 years. Scale-specific slopes become more negative from
$-0.032\;{}^\circ\mathrm C^{-1}$ at 126 km to
$-0.110\;{}^\circ\mathrm C^{-1}$ at 2,013 km; the corresponding negative-year
counts rise from 29 to 32. The negative association extends across the
within-month WBT distribution.

The estimated contrast depends on the definition of the spatial field.
Subtracting each site's 35-year monthly climatology reduces the estimate to
$-3.92\%$
($[-8.67,0.83]\%$); subtracting the site-year-month mean gives $-1.32\%$
($[-5.63,2.99]\%$). Site-month standardisation yields $-9.87\%$, but targets
a different dimensionless field. Together with the gradient decomposition,
these results show that high-day anomalies offset the
cross-site climatological gradient, while their own broad-scale energy changes
little.

The regime dates have a systematic but nonmonotone seasonal pattern. Averaged
over June--August, high days occur 1.96 days later than middle days (95\%
descriptive $(t)$-reference interval based on between-summer variation,
$[1.10,2.83]$ days): they occur 7.82 and 6.34 days
later in June and July, but 8.29 days earlier in August. Removing a separate
linear day-of-month trend within each site--year--month gives $-8.81\%$
($[-13.43,-4.19]\%$); subtracting the leave-one-year calendar-day climatology
gives $-9.81\%$ ($[-14.22,-5.40]\%$). The negative contrast persists after
both seasonal adjustments, although the weaker site-month and
site--year--month anomaly results show that different demeaning operations
target different spatial fields.

Calendar matching attenuates the contrast but preserves its direction. With
95\% descriptive $(t)$-reference intervals based on between-summer variation
in parentheses, the
$\pm3$-day estimator is $-5.21\%$ ($[-8.01,-2.41]\%$), with 24 of 33 summer
contrasts negative and 670 of 792 high days matched; the $\pm5$-day estimator is
$-5.72\%$ ($[-8.43,-3.02]\%$), with 25 negative summers and 750 matched high
days. Every one of the 99 month-year records has at least one eligible match
under both windows. The profiles run from $-2.32\%$ to $-8.87\%$ for
$\pm3$ days and from $-2.54\%$ to $-9.89\%$ for $\pm5$ days, retaining the
strengthening contraction with bandwidth.

\begin{figure}[t]
\centering
  \includegraphics[width=\textwidth, height=0.62\textheight, keepaspectratio]{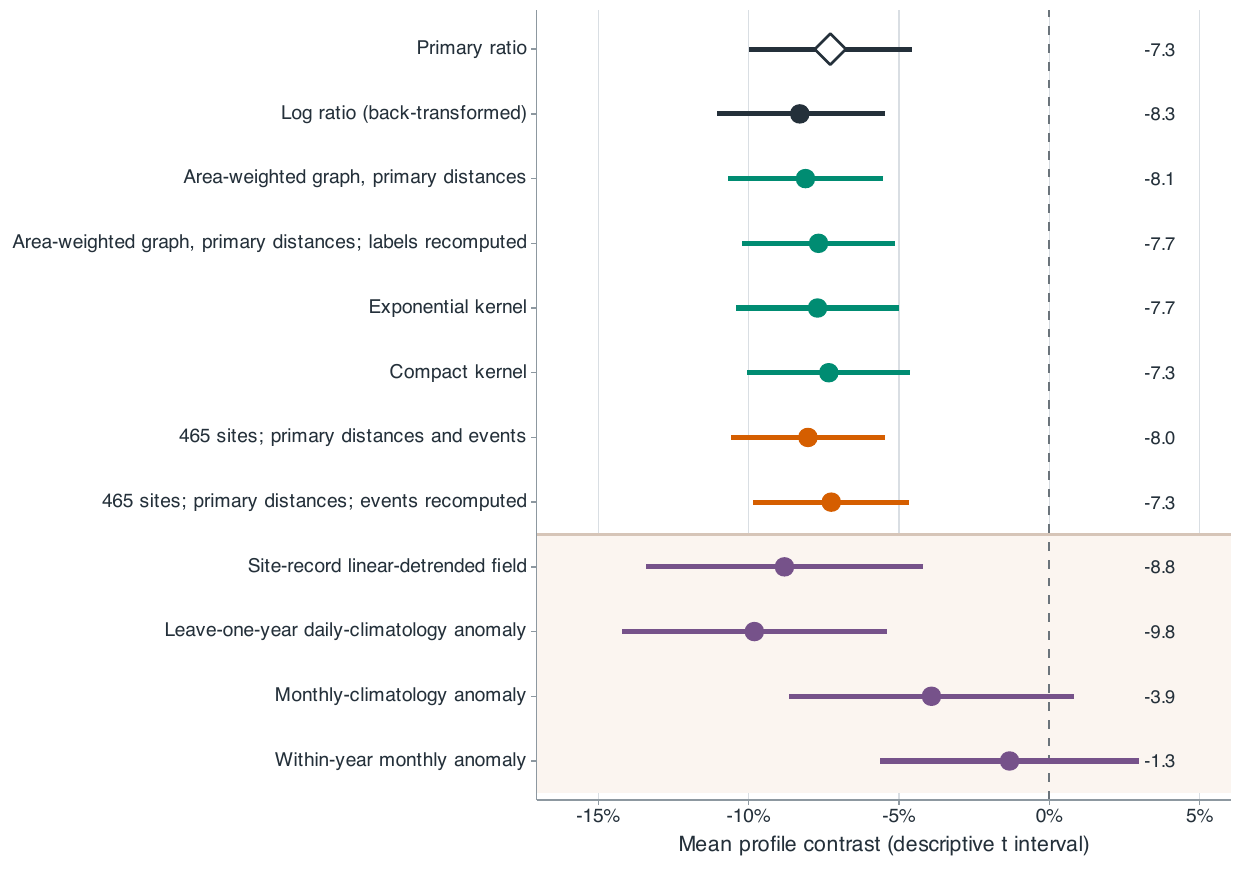}
  \caption{Robustness of the mean profile contrast. Dark rows denote the main
  and transformation analyses; green, orange and purple rows denote graph
  construction, spatial resolution and field definition. Points and bars are
  estimates and descriptive $t$-reference intervals based on between-summer
  variation.}
  \figalttext{A forest plot shows mostly overlapping negative
  estimates across transformation, graph and resolution choices. Two
  purple-shaded anomaly-field rows lie closest to zero.}
  \label{fig:application-robustness}
\end{figure}

All fixed-hour and daily-mean analyses yielded negative estimates. Using the
original event days, all 24 fixed-hour
estimates are negative, from $-10.07\%$ at 00 UTC to $-6.03\%$ at 11 UTC, and
daily-mean fields give $-7.58\%$. When labels are recomputed from each fixed
hour, all estimates remain negative, from $-14.65\%$ at 22 UTC to $-5.42\%$
at 11 UTC; labels recomputed from daily-mean WBT give $-9.14\%$. Peak hours
have nearly identical distributions in the original high and middle groups
(median 06 UTC). Figure~\ref{fig:application-robustness} compares the main
estimate with transformation, graph-construction, spatial-resolution and
field-definition alternatives.

The station comparison contains 175,172 exact-hour matches at 28
usable stations. ERA5-Land minus station WBT has a bias of
$-0.263\;{}^\circ$C, a mean absolute error of $0.938\;{}^\circ$C and a root
mean squared error of $1.270\;{}^\circ$C. The within-station centred
correlation is 0.916. The two offshore stations excluded from these summaries
had no finite values at their nearest ERA5-Land grid cells.

\begin{figure}[t]
\centering
  \includegraphics[width=\textwidth, height=0.62\textheight, keepaspectratio]{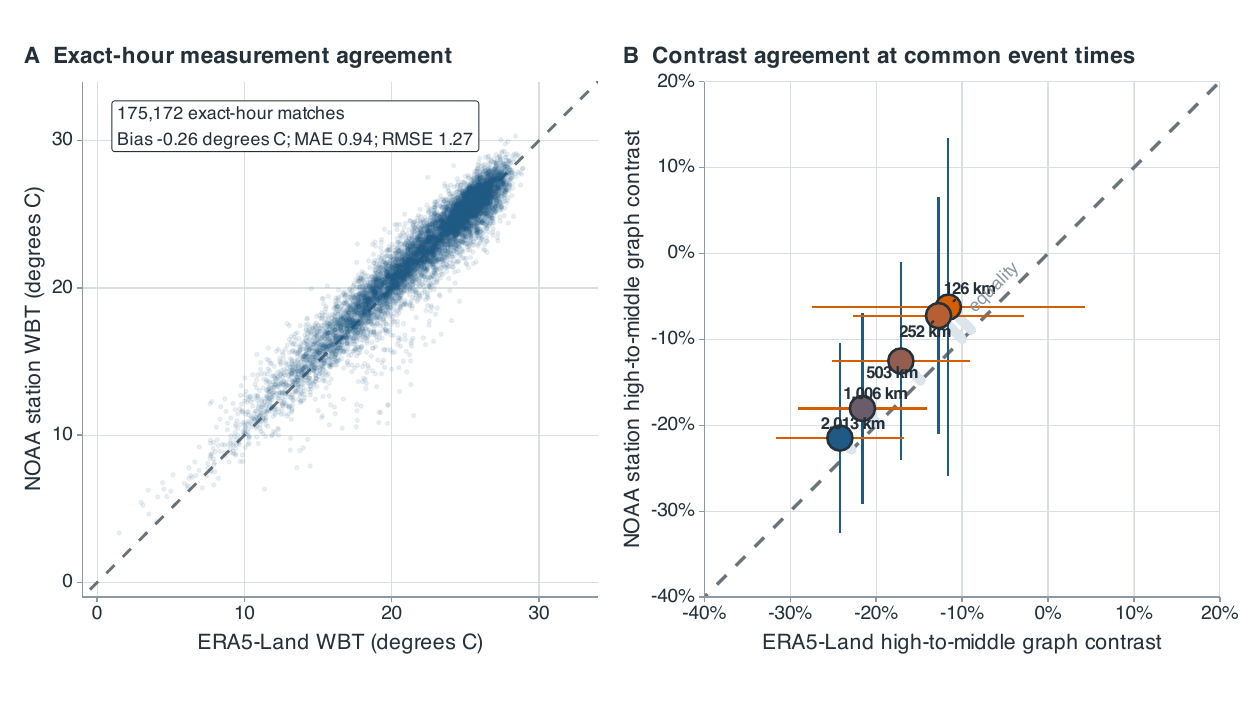}
  \caption{NOAA station and ERA5-Land agreement in ten station-comparison
  summers. (A) Exact-hour WBT at 28 stations, with summaries from
  all 175,172 matches. (B) High--middle graph contrasts at the five
  bandwidths, using ERA5-Land peak times and labels and identical station
  subsets. Dashed diagonals denote equality. Horizontal and vertical bars in
  panel B are paired descriptive $t$-reference intervals based on
  between-summer variation, not dependence-robust confidence intervals.}
  \figalttext{Two panels compare NOAA stations with
  ERA5-Land. Exact-hour WBT points in the left panel cluster around the
  equality line. In the right panel, station and ERA5-Land high--middle
  contrasts are negative at all five bandwidths and become more negative as
  the bandwidth increases; station estimates are slightly closer to zero.}
  \label{fig:noaa-validation}
\end{figure}

At Gaussian bandwidths 126, 252, 503, 1,006 and 2,013 km, the station
high--middle contrasts are $-6.23,-7.23,-12.50,-18.03$ and $-21.47\%$; the
corresponding ERA5-Land values on identical station subsets are
$-11.60,-12.74,-17.13,-21.60$ and $-24.23\%$. The numbers of negative station summers across the five scales are
6, 7, 8, 9 and 9. Averaged over the three predesignated
station-supported scales, the station contrast is $-17.34\%$
($[-28.49,-6.18]\%$), with 9 of 10 summer contrasts negative; the paired
ERA5-Land contrast is $-20.99\%$ ($[-28.57,-13.40]\%$), with all 10 negative.
The station and matched reanalysis estimates agree in direction at the three
designated broad scales. At 126 km, both intervals include zero and provide no
clear evidence of a local-scale contrast
(Figure~\ref{fig:noaa-validation}).

Of 240 scheduled high and 440 middle fields, 122 and 223, respectively, had
at least 10 stations. The high-minus-middle differences in mean station count
and eligible-field fraction were 0.16 sites (descriptive interval
$[-2.31,2.62]$) and 0.002 ($[-0.095,0.098]$). The three-band station contrast
was $-18.08\%$ on a fixed common support within each summer. Requiring at
least two, three or five retained days per regime gave $-13.42,-15.24$ and
$-14.56\%$; the five-day interval included zero because only six summers
remained. When regional means over the available station network defined both
the daily peak and event labels, the contrast was $-15.99\%$
($[-22.95,-9.02]\%$). Thus the broad direction survives the fixed-support
check and a station-based event definition on the available dynamic network,
but record sparsity limits precision.

Station elevations ranged from 3 to 2,063 m. In the WGS84 station calculation,
treating sea-level pressure as
surface pressure instead of applying the elevation conversion changed WBT by
0.072$^\circ$C on average in absolute value (maximum 1.346$^\circ$C) and the
three-band station contrast from $-17.37\%$ to $-17.29\%$.

\section{Discussion}

Across the 33 evaluation summers, the weighted squared-difference contrast was
$7.28\%$ lower when regional mean WBT was high; this corresponds to a
$3.7$--$4.0\%$ reduction on the RMS pairwise-difference scale. Twenty-eight
summer summaries were negative, and the protocol's three-component rule was
satisfied. The later product-shift result is exploratory and applies only
under the conditional product-invariance null. Contraction strengthened as
Gaussian bandwidth increased from $h=126$ to $h=2{,}013$ km, with the largest
estimated contraction under domain-wide graph support.
The 1950--1990 extension showed the same scale pattern, with a mean of
$-11.04\%$ and 40 of 41 summer contrasts below zero. Station data also agreed in
direction at 503 km and broader: the three-scale mean was $-17.34\%$, with
negative contrasts in 9 of 10 summers. At 126 km, the station and matched
reanalysis intervals both included zero. These contrasts describe spatial
arrangement, a feature distinct from the area exceeding a heat threshold.

Our analysis combines established scale-indexed graph dispersion,
finite-record randomisation and exact spatial accounting for an observational
field. The five graphs define a single finite-record summary while retaining its scale profile,
product shifts preserve within-record dependence among scales, and node and
energy decompositions recover the regional contrast exactly. Together, these
elements show how the change is distributed over nodes and scales. The
finite-record calculation and stochastic process model answer
different inferential questions. Product-shift rejection remained near
nominal in the simulated null settings. Separate simulations show why Student
and HAC intervals can be
unreliable with only 33 dependent summers.

The decompositions locate the contrast in the spatial field. After removing
each day's regional mean, high-day fields are relatively warmer in the north
and cooler in the south than middle-day fields, and the north--south component
accounts for most of the contrast. At broad scales, the
climatology--anomaly cross component accounts algebraically
for nearly all of the contraction, whereas the anomaly field's own energy
changes little. High regional mean WBT is therefore associated with a flatter
relative geographic pattern. The negative raw-field contrast persists under
area weighting, alternative kernels, fixed-hour analyses and calendar
matching. Its attenuation after site-month and site--year--month demeaning
further locates the result in the cross-site climatic gradient.

Several limitations remain. The 126-km bandwidth has the coarsest effective
spatial support and the weakest station evidence. WGS84 and boundary analyses
preserved direction and scale ordering, but estimates varied with the domain;
the inference therefore concerns the specified support rather than all of
eastern China.
The historical extension uses the same reanalysis product, and the 1950--1978
segment carries the documented preliminary-forcing caveat. The station
comparison covers 28 locations in ten summers and uses event times and labels
from ERA5-Land in the main comparison. Fixed-support and station-defined event
checks agree in direction, but stricter day-count requirements reduce the
record and widen uncertainty. The comparison measures external consistency,
not independent full-field replication. Pressure-level circulation is needed
to study a physical mechanism. Population-weighted fields, threshold areas
and health data are needed before drawing conclusions about exposure or
outcomes. Within the specified domain and record, high-regional-mean days had
weaker broad-scale WBT contrast while transient anomaly energy changed little.
Generalisation to a long-run climate process requires longer or independent
records.

\section*{Data availability}

ERA5-Land is available from the Copernicus Climate Data Store
(\url{https://cds.climate.copernicus.eu/datasets/reanalysis-era5-land}). The
NOAA ISD station archive \citep{smith2011} is available at
\url{https://www.ncei.noaa.gov/products/land-based-station/integrated-surface-database}.
The processed data, site and station manifests, quality-control records,
derived results and complete analysis code are available in the public GitHub
repository
\url{https://github.com/Stork343/eastern-china-wet-bulb-contraction}.
The submission-matched code, protocols, results and PDFs are frozen in the
release \url{https://github.com/Stork343/eastern-china-wet-bulb-contraction/releases/tag/jrssc-submission-v1-2026-08-13}.
The raw reanalysis and station archives remain available through their source
providers.

\section*{Acknowledgements}

OpenAI Codex was used for language checking and code debugging. The authors
reviewed all changes and take full responsibility for the manuscript and
analysis.

\section*{Funding}

This work was supported by the Beijing Natural Science Foundation, ``Theory,
Methodology and Applications of Functional Hierarchical Quantile Regression
Modeling'' (grant 1242005); the Fundamental Research Funds for the Central
Universities and the Research Funds of Renmin University of China, ``Robust
Statistical Inference for Complex Data'' (grant 25XNN015); and the Ministry of
Education Humanities and Social Sciences Research General Project, ``Research
on Spatial-temporal Quantile Regression Modeling: Theoretical Methods and
Applications'' (grant 25YJA910005).

\section*{Conflict of interest}

The authors declare no conflict of interest.

\end{document}